\documentclass{ws-procs9x6}

\def\sint{\int \!\!\!\!\!\!\!\! \sum_{p}}

\begin{document}

\title{QCD Phase Diagram \\ in Nonlocal Chiral Quark
Models}

\author{D. G\'omez Dumm}

\address{IFLP, CONICET $-$ Depto. de F\'{\i}sica,
Univ.\ Nac.\ de La Plata, \\
C.C. 67, 1900 La Plata, Argentina \\
E-mail: dumm@fisica.unlp.edu.ar}

\maketitle

\abstracts{The QCD phase diagram is analyzed within chiral quark models with
nonlocal separable interactions. For the case of two light flavors, we
describe in detail the characteristics of the chiral phase transition in the
$T-\mu$ plane, and analyze the color superconducting phase transition
taking place at relatively large chemical potentials. We also consider the
three-flavor version of the models, analyzing the predictions for masses and
decay constants of light pseudoscalar mesons.}

\section{Introduction}

The behavior of strongly interacting matter at finite temperature and/or
density is a subject of fundamental interest, with applications in
cosmology, astrophysics, and in the physics of relativistic heavy ion
collisions. In view of the difficulty found when dealing with strong
interactions in the nonperturbative regime, it has not been possible yet
to obtain detailed information about the full $T-\mu$ phase diagram
directly from QCD, and most of the theoretical work on the subject relies
in low energy effective models. One of the most popular ones is that
proposed long ago by Nambu and Jona-Lasinio~(NJL)\cite{NJL}, in which
quarks interact through local four fermion couplings. In the last few
years, covariant nonlocal extensions of this model have been
proposed\cite{Rip97}, mainly motivated by some approaches to low energy
quark dynamics (instanton liquid model, Schwinger-Dyson resummation
techniques) in which nonlocality arises in a natural way. It has also been
shown that these nonlocal theories present several advantages in
comparison with the standard NJL scheme. Here we analyze, in this context,
the structure of the QCD phase diagram. First, we consider the case of
two-flavor models, studying the features of the chiral phase transition
for finite $T$ and $\mu$ and the presence of a possible superconducting
phase in the region of large $\mu$. Then we also address the three-flavor
case, studying the predicted pattern of pseudoscalar meson masses and
decay constants. The analysis summarized here is developed in detail in
Refs.\cite{nos1,nos2,gru,nos3}.

\section{Formulation}

We start by quoting the Euclidean quark effective action
\begin{equation}
\! S_E = \int d^4 x \left\{ \bar \psi (x) \left(- i \rlap/\partial  +
m_c \right) \psi (x) - \frac{G}{2} [ j^S(x) j^S(x) + j_k^P(x) j_k^P(x) ]
\right\},\!\!\!
\label{se}
\end{equation}
where $\psi\equiv (u\ d)^T$, while the fermion currents $j^S(x)$ and
$j_k^P(x)$ are given by
\begin{equation} \left\{ \begin{array}{c}
j^S(x) \\ \rule{0cm}{0.35cm} j_k^P(x) \end{array} \right\} = \int d^4 y\
d^4 z \  r(y-x) \ r(x-z) \  \bar \psi(y) \
\left\{\!\!\begin{array}{c}
\leavevmode\hbox{\small1\kern-3.8pt\normalsize1}
\\ i\gamma_5\tau_k \end{array}
\!\!\right\} \ \psi(z) \ .
\end{equation}
Here $r(x-y)$ is a soft covariant nonlocal regulator, which in the case of
the instanton liquid model has a definite form in terms of modified Bessel
functions. In the following we will consider the relatively simple case of
a Gaussian regulator $r(p^2) = \exp (-p^2/2\Lambda^2)$, where $\Lambda$ is
a free parameter of the model, playing the r\^ole of an ultraviolet cutoff
scale. In any case, our results are found to be qualitatively similar for
other regulator shapes. Notice that the constituent quark mass $m_c$ (we
assume isospin symmetry) and the coupling constant $G$ are also free
parameters of the model.

We perform now a bosonization of the partition function of the theory, $Z
= \int D\bar \psi D\psi \exp(-S_E)$. To do this we introduce scalar and
pseudoscalar meson fields $\sigma(x)$, $\pi_k(x)$, integrating out the
quark degrees of freedom. Then we carry out the mean field approximation
(MFA), in which the meson fields are expanded around their translational
invariant vacuum expectation values $\bar\sigma$ and $\bar\pi_k$ (the
latter vanish due to charge and isospin symmetries). The extension to
finite $T$ and $\mu$ is obtained by performing the replacement
\begin{equation}
\int \frac{d^4 p}{(2\pi)^4}\ F(p_4,\vec p) \ \to \
\sint F(p_4,\vec p) \equiv
T \sum_{n=-\infty}^\infty \int \frac{d^3 p}{(2\pi)^3}\;
F(\omega_n - i\mu,\vec p)
\end{equation}
in the four-momentum integrals arising from the fermion determinant (here
$\omega_n$ are Matsubara frequencies corresponding to fermion
modes). The grand canonical thermodinamical potential per unit volume is
then given by
\begin{equation}
\omega_{\mbox{\tiny MFA}} (T,\mu,m_c) = -\,\frac{T}{V}\log Z_{\mbox{\tiny
MFA}} = \frac{\bar \sigma^2}{2 G} - 4 N_c \; \sint \log \left[ p^2 +
\Sigma^2(p^2) \right] ,
\end{equation}
where $\Sigma(p^2) = m_c + \bar \sigma\; r^2(p^2)$ is the constituent
(momentum-dependent) quark mass. Now $\bar\sigma(T,\mu,m_c)$ can be
obtained from $\partial\omega_{\mbox{\tiny MFA}}/\partial\bar\sigma=0$
(gap equation), and one can easily derive the expressions for other
relevant physical quantities, such as the quark condensate $\langle\bar q
q\rangle$, quark density $\rho$, chiral susceptibility $\chi_{V,T,\mu} = -
(\partial\langle\bar q q\rangle/\partial m_c)_{T,\mu}$, pressure, energy
density, etc.

\section{$\!$Chiral phase transition: analytical and numerical analysis}

In order to study the physical quantities, the Matsubara sums can be
worked out using complex plane analysis. While in the standard (local) NJL
model the Euclidean quark propagator has just two simple poles at $p_4=\pm
i\sqrt{\vec p^2+M^2}$, being $M$ the quark constituent mass, in the case
of nonlocal models the regulator leads in general to a complicated
structure of poles and cuts in the complex plane. Depending on the
parameters of the model and the shape of the regulator, there may be some
poles on the real axis or not; it has been suggested that this last
situation is a sort of realization of confinement, in the sense that the
effective quark propagator has no poles at real energies\cite{BB95}, and
one can talk about ``confining'' and ``nonconfining'' parameter sets. From
our analysis it is seen\cite{nos1,nos2} that using adequate integration
paths, and introducing some auxiliary complex occupation number functions,
the Matsubara sums can be converted into integrals. Moreover, we have
found that in all considered cases the behavior of the physical quantities
is strongly dominated by the contribution of the first pole of the quark
propagator.

We have carried out numerical calculations to evaluate the dependence of
the mentioned physical quantities with $T$ and $\mu$. To proceed, we have
adjusted the free parameters $\Lambda$, $m_c$ and $G$ so as to reproduce
the empirical values of the pion mass and decay constant, whereas as the
third input we have chosen some fixed values of $\Sigma(0)$. Both
confining and nonconfining parameter sets have been considered, studying
in each case the features of the chiral phase transition. In general, the
results have been found to be qualitatively similar (a detailed comparison
can be found in Refs.\cite{nos1,nos2}); for definiteness we present here
the curves corresponding to one of these parameter sets, namely
$\Lambda=760$~MeV, $m_c=7.7$~MeV and $G=30$~GeV$^{-2}$.

\begin{figure}[ht]
\centerline{\epsfxsize=4.5in\epsfbox{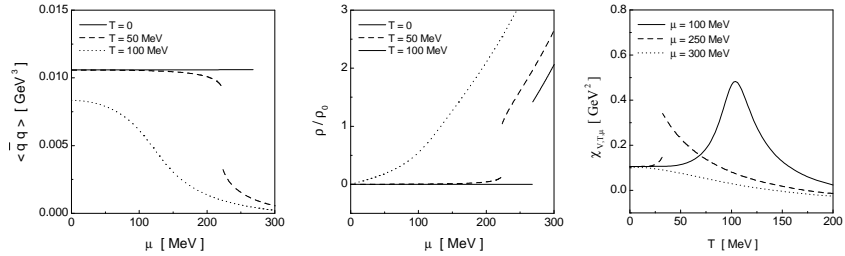}}
\caption{Behavior of the chiral condensate (left), the density (center) and
the chiral susceptibility (right) with the temperature and the chemical
potential.
\label{fig1}}
\end{figure}

\begin{figure}[ht]
\centerline{\epsfxsize=3.2in\epsfbox{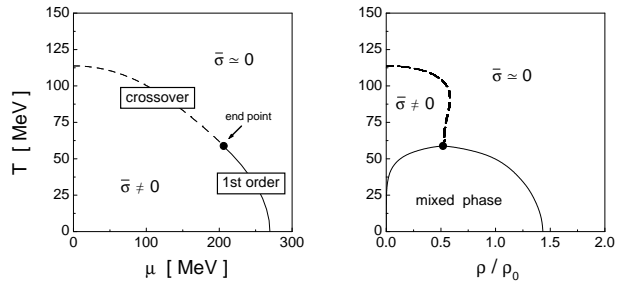}}
\caption{Phase transition temperature as a function of the
chemical potential (left) and the density (right). Solid (dashed) lines
stand for first order transition (crossover).
\label{fig2}}
\end{figure}

Our main results are shown in Fig.~\ref{fig1}, where we quote the values
of the quark condensate, the density (relative to nuclear matter density
$\rho_0$) and the chiral susceptibility as functions of $T$ and $\mu$. By
looking at the curves for the quark condensate, it is seen that for $T =
0$ there is a first order phase transition at a given critical chemical
potential $\mu_c$. The latter goes down as $T$ is increased, until one
reaches an end point where the first order phase transition turns into a
smooth crossover. For this parameter set the end point is located at
$(T_E,\mu_E)=(55\ {\rm MeV},210\ {\rm MeV})$. In the crossover region, the
transition is also well defined by the presence of marked peaks in the
chiral susceptibility, the sharpness of these peaks giving a measure of
the crossover steepness. Thus we arrive at the phase diagrams shown in
Fig.~\ref{fig2}, in which we plot the phase transition temperature as a
function of the chemical potential (left) and the density (right). The
latter includes a ``mixed phase'' zone, which can be interpreted as a
region where droplets containing light quarks of mass $m_c$ coexist with a
gas of constituent, massive quarks. It should be noticed that the model
predicts a critical temperature at $\mu=0$ of about 100 MeV, somewhat
below the results obtained from lattice calculations which suggest this
value to be $\sim 140$ $-$ 190 MeV.

\section{Extensions to diquark couplings and three flavors}

The ability of this kind of models to allow for a two-flavor
superconducting (2SC) phase at large chemical potentials is also under
study\cite{gru}. This analysis considers an effective quark-quark
interaction in the Euclidean action (\ref{se}), carrying a new coupling
constant $h$. Preliminary results show that for low $T$ a first order
phase transition into a 2SC phase is expected at some critical value of
$\mu$. A similar behavior is found when $T$ is increased, until one
reaches a ``triple point'' in which all three phases (hadronic, chiral and
2SC) coexist. The phase diagram is then completed with the inclusion of a
second order phase transition line between the chiral and the 2SC phases.
Unfortunately, the precise features of the phase diagram depend on the
value of the diquark coupling $h$, which cannot be fixed from low energy
phenomenology. A detailed discussion in this sense is given in
Ref.\cite{gru}.

Finally, we have considered\cite{nos3} the inclusion of strange degrees of
freedom, extending the flavor symmetry of the model to $SU(3)$ and
incorporating the the $U(1)_A$ symmetry breaking through a nonlocal 't
Hooft-like coupling. As a first step we have concentrated on the
evaluation of meson masses and decay constants, finding our results in
quite good agreement with empirical values, in particular in the case of
the ratio $f_K/f_\pi$ and the anomalous decay $\pi^0\to\gamma\gamma$
(which are problematic in the standard NJL model). We have also obtained a
reasonable description of the $\eta-\eta'$ phenomenology. Remarkably, this
requires the presence of two significantly different state mixing angles.

Clearly, the two-flavor diquark channel analysis cannot be trusted beyond
the strange quark mass threshold, and the $SU(3)$ version of the model
should be addressed, leading presumably to the presence of a color flavor
locking phase. We expect to report on these issues in forthcoming works.

\section*{Acknowledgments}
I am indebted to N.N.~Scoccola and A.~Grunfeld for useful discussions, and
to the organizers of SEWM04 for financial aid. This work has been
partially supported by ANPCyT (Argentina) under grant PICT02-03-10718.

\end{document}